\documentclass{eptcs}
\usepackage{graphicx}
\usepackage{todonotes}
\usepackage{breakurl}
\usepackage{paralist}

\usepackage{color}
\definecolor{lstbg}{rgb}{0.98,0.98,0.98}

\usepackage{listings}
\lstdefinelanguage{gretl}{
  morekeywords={E, V, as, bag, eSubgraph, end, exists!, exists, false, forall,
    from, import, in, let, list, null, path, pathSystem, rec, report,
    reportBag, reportMap, reportSet, thisEdge, thisVertex, map,
    role,    
    set, store,
    true, tup, using, vSubgraph, where, with, to},
  morekeywords={new, protected, public, static, final, private, if, switch,
    for, break, default, throw, void, else, return, continue},
  morekeywords={[2]and, avg, contains, containsKey, count, degree, depth,
    difference, distance, dividedBy, edgesConnected, edgesFrom, edgesTo,
    edgeTrace, edgeTypeSet, endVertex, enumConstant, equals, extractPath,
    getEdge, get, getValue, getVertex, grEqual, grThan, hasAttribute, hasType,
    id, inDegree, innerNodes, intersection, isAcyclic, isEmpty, isA, isCycle,
    isIn, isIsolated, isLoop, isNeighbour, isNull, isParallel, isReachable,
    isSibling, isSubPathOfPath, isSubSet, isSuperSet, isTrail, isTree, keySet,
    leaves, leEqual, leThan, matches, maxPathLength, minPathLength, minus,
    modulo, nequals, nodeTrace, not, nthElement, theElement, or, outDegree,
    package-info, parent, pathConcat, pathLength, pathSystem, plus, pos,
    reachableVertices, recordInstance, reMatch, schemaFunctions, siblings,
    squareRoot, startVertex, subtypes, sum, supertypes, symDifference, times,
    toString, type, typeName, typeSet, uminus, union, values,
    vertexTypeSet, weight, xor, error},
  emph={AddMappings, Assert, CreateSubgraph, CreateVertexClass,
    CreateAbstractVertexClass, CreateEdgeClass, CreateAbstractEdgeClass,
    AddSubClass, AddSubClasses, AddSuperClass, AddSuperClasses, CopyDomain,
    CreateAttribute, CreateAttributes, CreateListDomain, CreateSetDomain,
    CreateMapDomain, CreateRecordDomain, CreateEnumDomain, CreateVertices,
    CreateEdges, ExecuteTransformation, MatchReplace, MergeVertices, Delete,
    SetAttributes, RedefineFromRole, RedefineFromRoles, RedefineToRole,
    RedefineToRoles, transformation, Iteratively},
  emphstyle=\bf\underbar,
  sensitive=true,
  string=[b]{\\"},
  morecomment=[l]{//}}

\makeatletter
\lst@AddToHook{TextStyle}{\let\lst@basicstyle\footnotesize\sffamily}
\makeatother

\newcommand{\myemail}[0]{\email{horn@uni-koblenz.de}}
\lstdefinestyle{sm}{language=gretl, name={sm}, firstnumber=auto}

\def\titlerunning{Solving the TTC 2011 Reengineering Case with GReTL}
\def\authorrunning{Tassilo Horn}

\title{Solving the TTC 2011 Reengineering Case with GReTL}
\author{Dipl.-Inform. Tassilo Horn\\
  \myemail\\
  Institute for Software Technology\\
  University Koblenz-Landau, Campus Koblenz}

\begin{document}
\maketitle
\begin{abstract}
  This paper discusses the GReTL reference solution of the TTC 2011
  Reengineering case \cite{programunderstandingcase}.  Given a Java syntax
  graph, a simple state machine model has to be extracted.  The submitted
  solution covers both the \emph{core task} and the two \emph{extension tasks}.
\end{abstract}

\section{Introduction}
\label{sec:introduction}

GReTL (Graph Repository Transformation Language, \cite{gretl-icmt2011}) is the
operational transformation language of the TGraph technological space
\cite{tgapproach08}.  Models are represented as typed, directed, ordered, and
attributed graphs.  GReTL uses GReQL (Graph Repository Query Language,
\cite{FestschriftNagl2010}) for its querying part.

In contrast to most other transformation languages, GReTL transformations
usually construct the target metamodel (\emph{schema}) on their own, thereby
specifying one graph conforming to this new schema as target graph of the
transformation.  For this purpose, GReTL provides a slim set of four
transformation operations, which are derived from the metametamodel of the
TGraph technological space (the \textit{GraphUML metaschema}).  There is an
operation \emph{CreateVertexClass}, which creates a new node type
(\emph{VertexClass}) in the target schema and a set of vertices of this new
type in the target graph.  Likewise, there is an operation
\emph{CreateEdgeClass}, which creates a new edge type (\emph{EdgeClass}) in the
target schema and a set of edges of this new type in the target graph.  Since
schemas allow for multiple inheritance between vertex as well as edge classes,
there is an operation \emph{AddSubClass} to create specialization relationships
in the target schema.  Finally, there is an operation \emph{CreateAttribute},
which creates a new attribute for a vertex or edge class and which assigns
values to the elements for which that new attribute is defined.  The vertices
and edges that have to be created in the target graph as well as the function
assigning values are specified in terms of queries on the transformation's
source graph.

\section{Task Solutions}
\label{sec:task-solutions}

In this section, all tasks are discussed in sequence, and the GReTL operations
and GReQL queries are explained when they come along.  The solution can be run
on the SHARE image \cite{share-demo}.

\subsection{The Core Task}
\label{sec:core-task}

The core task is responsible for creating \textsf{States} and
\textsf{Transitions} without setting the attributes of the latter.  Because the
JaMoPP metamodel \cite{jamopp09} splits its types into various packages, we
import the packages from which elements are used, so that we can refer to these
types without having to qualify them.

\begin{lstlisting}[style=sm]
import classifiers.*; import types.*;      import members.*;
import references.*;  import statements.*; import modifiers.*;
\end{lstlisting}

The first task is the creation of \textsf{State} vertices.  As explained in the
task description, there is an abstract Java class named \textsf{State}, and all
concrete subclasses can be considered states.  At first, we bind abstract
\textsf{State} class to a variable \textsf{abstractStateClass}, so that we can
easily refer to it in the transformation.

\begin{lstlisting}[style=sm]
abstractStateClass := theElement(from c: V{Class}
                                 with c.name = "State" reportSet c end);
\end{lstlisting}

The \emph{from-with-reportSet} expression calculates the set of classes which
are named ``State''.  The function \textsf{theElement()} extracts the single
element of a collection consisting of only one element and throws an exception
if the collection's size is not one.  This expresses the assumption that there
is exactly one state class.  Finally, this class is assigned to the variable
\textsf{abstractStateClass}.

The first transformation operation invoked is \textsf{CreateVertexClass}.  It
creates a new vertex class \textsf{State} in the target schema.  The query
following the arrow symbol is evaluated on the source graph and has to result
in a set.  For each member of this set (\emph{archetype}), a new vertex
(\emph{image}) of the just created type is instantiated in the target graph.
The mappings from archetypes to target graph images are automatically saved in
a function corresponding to the target metamodel vertex class ($img_{State}$).

\begin{lstlisting}[style=sm]
CreateVertexClass State
  <== from c: {Class} & (<>--{extends} <>--{classifierReferences} -->{target})+
                         abstractStateClass
      with isEmpty(c <>--{annotationsAndModifiers} & {Abstract})
      reportSet c end;
\end{lstlisting}

The query specifies a set of \textsf{Class} vertices.  The variable \textsf{c}
iterates over \textsf{Class} vertices, for which a path to the vertex
\textsf{abstractStateClass} exists.  The structure of this path is specified
using a \emph{regular path expression} \cite{FestschriftNagl2010}.  First, a
containment edge with role name \textsf{extends} at the far end has to be
traversed, followed by another containment edge with role name
\textsf{classifierReferences}, followed by a forward edge with role name
\textsf{target}.  This is exactly how subclasses relate to their superclass.
The \textsf{+} specifies a one-or-many iteration.  Thus, \textsf{c} is bound
not only to direct subclasses of \textsf{abstractStateClass}, but also to
indirect ones.  The \textsf{with} part ensures that \textsf{c} is not abstract,
i.e., it must not reference an \textsf{Abstract} vertex using an edge with
containment semantics and far end role name \textsf{annotationsAndModifiers}.
For any non-abstract class that extends the abstract state class either
directly or indirectly, a new target graph \textsf{State} vertex is created.
The mappings from classes to states are stored in a function
\textsf{img\_State}, which can be used in following operation calls for
navigating between archetypes and images.

The next operation creates the \textsf{name} attribute of type \textsf{String}
for the \textsf{State} vertex class, and it sets the attribute values for the
vertices created by the last operation call.

\begin{lstlisting}[style=sm]
CreateAttribute State.name : String
  <== from c: keySet(img_State) reportMap c -> c.name end;
\end{lstlisting}

The query of the \textsf{CreateAttribute} operation has to result in a map
assigning values of the attribute's type to archetyps.  Here, the map assigns
class names to \textsf{State} archetypes, so the state names are set to the
names of the classes they were created for.

The \textsf{CreateEdgeClass} operation is used to create a new edge type
\textsf{Transition} in the target metamodel defined between \textsf{State}
vertices with the given role names and default multiplicities (0,*).  The query
has to result in a set of triples.  In each triple, the first component
specifies the archetype of the new edge to be created.  The second and third
component specify the archetypes of the start and end vertices.  For each
archetype, a new edge of the just created type is created in the target graph,
starting at the vertex that is the image of the second component and ending at
the vertex that is image of the third component.

\begin{lstlisting}[style=sm]
CreateEdgeClass Transition from State role src to State role dst
  <== from c1, c2: keySet(img_State),
           callingMethod: c1 <>--{members} & {Method},
           call: callingMethod <>--+ & {MethodCall}
      with call -->{target} instanceMethod
        and not isEmpty(call <>--{next} & {MethodCall} -->{target}
                        & {Method @ thisVertex.name = "activate"})
      reportSet tup(c1, callingMethod, c2, instanceMethod), c1, c2 end
      where instanceMethod := theElement(c2 <>--{members}
                                & {Method @ thisVertex.name = "Instance"});
\end{lstlisting}

In the query, \textsf{c1} and \textsf{c2} iterate over \textsf{State}
archetypes, i.e., source graph \textsf{Class} vertices extending the abstract
state class.  The variable \textsf{callingMethod} is bound to all methods of
the class bound to \textsf{c1} one after the other.  In turn, \textsf{call} is
bound to every \textsf{MethodCall} occuring somewhere in
\textsf{callingMethod}'s body using the regular path expression
\lstinline{<>--+}, which calculates all method call vertices reachable from
\textsf{callingMethod} by traversing edges with containment semantics one or
many times.  The variable \textsf{instanceMethod} is bound to the singleton
\textsf{Instance()} \textsf{Method} of \textsf{c2} using a \textsf{where}
binding.  The predicates in the \textsf{with} part ensurne that the method call
\textsf{call} indeed invokes the \textsf{instanceMethod} and that on the object
returned by the call, the \textsf{activate()} method is invoked.  For each
variable combination fulfilling the predicates, a triple is reported.  The
first component, i.e., the archetype for a new \textsf{Transition} edge, is a
tuple containing the currently active state class \textsf{c1}, its method that
contains the activation call (\textsf{callingMethod}), the new state class
\textsf{c2}, and its \textsf{instanceMethod}.  The archetype of the new
transition's start state is \textsf{c1} and the transition leads to the state
that is the image of \textsf{c2}.

This is all the core task requires.  In the remainder, the extension solutions
are discussed.

\subsection{Extension 1: Triggers}
\label{sec:ext-triggers}

With respect to triggers, four cases have to be distinguished:
\begin{inparaenum}[(1)]
\item If the transition occurs in a method except for \textsf{run()}, then that
  method's name is the trigger.
\item If it occurs in a \textsf{switch} statement in the \textsf{run()} method,
  then the corresponding \textsf{case}'s enum constant name is the trigger.
\item If it occurs in a \textsf{catch} block, then the caught exception's type
  name is the trigger.
\item Else, the trigger should be set to the string \lstinline{``--''}.
\end{inparaenum}
The four situations are covered by different operation calls.

\subparagraph{Non-\textsf{run()} methods and default value.}

The \textsf{CreateAttribute} operation is used to create the \textsf{trigger}
attribute of type \textsf{String} for the \textsf{Transition} edge class.  The
string \lstinline{``--''} is chosen as default value, which handles the fourth
case above.

\begin{lstlisting}[style=sm]
CreateAttribute Transition.trigger : String = '"--"'
  <== from t: keySet(img_Transition)
      with t[1].name <> "run"
      reportMap t -> t[1].name end;
\end{lstlisting}

The query returns a map that assigns to every \textsf{Transition} archetype
whose second component's \textsf{name} is not \textsf{``run''} the value of its
\textsf{name}.  When looking at the \textsf{CreateEdgeClass} call for
\textsf{Transition} in the core task, the second component of
\textsf{Transition} archetypes (\textsf{t[1]}) is exactly the method in which
the activation of the new state occured.

\subparagraph{Switch statements.}

Because the \textsf{trigger} attribute has already been created by the previous
\textsf{CreateAttribute} call, the \textsf{SetAttributes} operation is used,
which only works on the instance level and requires an existing attribute given
by its qualified name (\textsf{Transition.trigger}).

\begin{lstlisting}[style=sm]
SetAttributes Transition.trigger
  <== from t: keySet(img_Transition),
           case: t[1] <>--+ & {Switch}<>--{cases},
           cond: case <>--{condition} -->{target} & {EnumConstant}
      with t[1].name = "run" and case <>--+ & {MethodCall} -->{target} t[3]
      reportMap t -> cond.name end;
\end{lstlisting}

The query iterates over \textsf{Transition} archetypes (4-tuples) using the
variable \textsf{t}.  \textsf{case} is bound to all \textsf{Case} vertices
reachable by diving into the body of the activating method referenced by
\textsf{t[1]}, reaching a \textsf{Switch} vertex, and selecting its
\textsf{Case} vertices one after the other.  In turn, \textsf{cond} is bound to
the \textsf{EnumConstant} that is the condition of the \textsf{case}.  The
predicates in the \textsf{with} part ensure that the activation of the next
state occurs inside the \textsf{run()} method, and that the call of the
\textsf{Instance()} method (the fourth component in the transition archetype
tuples) indeed occurs in the body of \textsf{case}.  The map assigns the names
of the \textsf{EnumConstant} used in the case of the switch statement to
\textsf{Transition} archetypes.

\subparagraph{Catch blocks.}

Again, the \textsf{SetAttributes} operation is used.

\begin{lstlisting}[style=sm]
SetAttributes Transition.trigger
  <== from t: keySet(img_Transition), catch: t[1] <>--+ & {CatchBlock}
      with t[1].name = "run" and catch <>--+ & {MethodCall} -->{target} t[3]
      reportMap t -> theElement(catch -->{parameter} -->{typeReference}
                                  -->{classifierReferences} -->{target}).name end;
\end{lstlisting}

The query iterates over \textsf{Transition} archetypes using the variable
\textsf{t}.  From the method containing the activation call (\textsf{t[1]}),
all \textsf{CatchBlock} vertices contained in its body are iterated.  The
\textsf{with} part ensures the activation is inside the \textsf{run()} method
and the activation call occurs inside the \textsf{catch} block.  For all
combinations of \textsf{t} and \textsf{catch} where the predicates hold, the
name of the caught exception is the trigger.

\subsection{Extension 2: Actions}
\label{sec:ext2-actions}

The second extension task deals with setting the \textsf{action} attribute of
\textsf{Transition} edges.  The value of this attribute is the name of the
enumeration constant provided as argument to a \textsf{send()} method call
appearing in the same block as the activation of the next state.  If there is
no such call, the attribute should be set to \lstinline{``--''}.  The
\textsf{CreateAttribute} operation is used to create the new \textsf{action}
attribute of type \textsf{String} for the edge class \textsf{Transition} with
the default value \lstinline{``--''}.

\begin{lstlisting}[style=sm]
CreateAttribute Transition.action : String = '"--"'
  <== from t: keySet(img_Transition),
           container: t[1] <>--* & {StatementListContainer},
           sendCall: container <>--{statements} <>--{expression} & {MethodCall}
      with not isEmpty(sendCall -->{target} & {Method @ thisVertex.name = "send"})
        and container <>--{statements} <>--{expression}
                      -->{next}* & {MethodCall}-->{target} t[3]
      reportMap t -> theElement(sendCall <>--{arguments}
                                         <>--{next} -->{target}).name end;
\end{lstlisting}

The query iterates over the \textsf{Transition} archetype tuples \textsf{t}.
The variable \textsf{container} is bound to all blocks
(\textsf{StatementListContainer}) contained in the method that contains the
activation call of the next state one after the other, i.e., first it is bound
to the method body, then to blocks of \textsf{if}, \textsf{catch}, or
\textsf{catch} statements.  The variable \textsf{sendCall} is in turn bound to
all \textsf{MethodCall} vertices contained in the \textsf{container}.  The
first predicate in the \textsf{with} part of the query ensures that
\textsf{sendCall} invokes in fact a \textsf{Method} with \textsf{name}
\textsf{``send''}, and the second predicate ensures that in the block
\textsf{container} the activation of the next state occurs.  The query reports
a map which assigns the name of the argument given to the \textsf{send()}
method in \textsf{sendCall} to the selected \textsf{Transition} archetypes.

This was the last operation of the \texttt{ExtractStateMachines.gretl}
transformation.  The complete task could be solved with only 45 lines of
transformation source code.

\section{Conclusion}
\label{sec:conclusion}

In this paper, the complete GReTL reference solution of the program
understanding case has been discussed in details.  In this conclusion, some
statements about the evaluation criteria are made.

The solution covers the core as well as both extension tasks, so it is
\emph{complete}.  The \emph{correctness} of the solution has been validated
with the three provided input models.  Transforming them always results in the
same target model, except for the (for this case irrelevant) order of elements
in the models.

With respect to \emph{conciseness}, only 45 lines of transformation source code
for this non-trivial task is very good.

The \emph{performance} of the solution is also good.  On SHARE
\cite{share-demo}, all three input models can be transformed in less than two
seconds.

With respect to \emph{understandability}, one may argue that GReTL's conception
of incrementally constructing the target schema and graph simultaneously, its
traceability concept of archetypes and images, and the query language GReQL are
not easy to grasp at first.  However, they have proven being flexible and
expressive, so it might be worth the initial steep learning curve.

\bibliographystyle{eptcs}
\bibliography{bibliography}

\begin{thebibliography}{1}
\providecommand{\bibitemdeclare}[2]{}
\providecommand{\urlprefix}{Available at }
\providecommand{\url}[1]{\texttt{#1}}
\providecommand{\href}[2]{\texttt{#2}}
\providecommand{\urlalt}[2]{\href{#1}{#2}}
\providecommand{\doi}[1]{doi:\urlalt{http://dx.doi.org/#1}{#1}}
\providecommand{\bibinfo}[2]{#2}

\bibitemdeclare{incollection}{FestschriftNagl2010}
\bibitem{FestschriftNagl2010}
\bibinfo{author}{J\"urgen Ebert} \& \bibinfo{author}{Daniel Bildhauer}
  (\bibinfo{year}{2010}): \emph{\bibinfo{title}{{Reverse Engineering Using
  Graph Queries}}}.
\newblock In: {\sl \bibinfo{booktitle}{Graph Transformations and Model Driven
  Engineering}}, \bibinfo{series}{LNCS 5765}, \bibinfo{publisher}{Springer},
  pp. \bibinfo{pages}{335--362}, \doi{10.1007/978-3-642-17322-6\_15}.

\bibitemdeclare{inproceedings}{tgapproach08}
\bibitem{tgapproach08}
\bibinfo{author}{J\"urgen Ebert}, \bibinfo{author}{Volker Riediger} \&
  \bibinfo{author}{Andreas Winter} (\bibinfo{year}{2008}):
  \emph{\bibinfo{title}{{G}raph {T}echnology in {R}everse {E}ngineering, {T}he
  {T}{G}raph {A}pproach}}.
\newblock In \bibinfo{editor}{R.~Gimnich}, \bibinfo{editor}{U.~Kaiser},
  \bibinfo{editor}{J.~Quante} \& \bibinfo{editor}{A.~Winter}, editors: {\sl
  \bibinfo{booktitle}{10th {W}orkshop {S}oftware {R}eengineering ({W}{S}{R}
  2008)}}, {\sl \bibinfo{series}{GI Lecture Notes in Informatics}}
  \bibinfo{volume}{126}, \bibinfo{publisher}{GI}, pp. \bibinfo{pages}{67--81}.

\bibitemdeclare{incollection}{jamopp09}
\bibitem{jamopp09}
\bibinfo{author}{Florian Heidenreich}, \bibinfo{author}{Jendrik Johannes},
  \bibinfo{author}{Mirko Seifert} \& \bibinfo{author}{Christian Wende}
  (\bibinfo{year}{2010}): \emph{\bibinfo{title}{{Closing the Gap between
  Modelling and Java}}}.
\newblock In \bibinfo{editor}{Mark van~den Brand}, \bibinfo{editor}{Dragan
  Gasevic} \& \bibinfo{editor}{Jeff Gray}, editors: {\sl
  \bibinfo{booktitle}{Software Language Engineering}}, {\sl
  \bibinfo{series}{Lecture Notes in Computer Science}} \bibinfo{volume}{5969},
  \bibinfo{publisher}{Springer}, pp. \bibinfo{pages}{374--383},
  \doi{10.1007/978-3-642-12107-4\_25}.

\bibitemdeclare{misc}{share-demo}
\bibitem{share-demo}
\bibinfo{author}{Tassilo Horn}: \emph{\bibinfo{title}{{SHARE} demo related to
  the paper {Solving the TTC 2011 Reengineering Case with GReTL}}}.
\newblock
  \bibinfo{howpublished}{\url{http://is.ieis.tue.nl/staff/pvgorp/share/?page=C%
onfigureNewSession&vdi=Ubuntu_10.04_TTC11_gretl-cases.vdi}}.

\bibitemdeclare{inproceedings}{programunderstandingcase}
\bibitem{programunderstandingcase}
\bibinfo{author}{Tassilo Horn} (\bibinfo{year}{2011}):
  \emph{\bibinfo{title}{Program Understanding: A Reengineering Case for the
  Transformation Tool Contest}}.
\newblock In \bibinfo{editor}{Pieter {Van Gorp}}, \bibinfo{editor}{Steffen
  Mazanek} \& \bibinfo{editor}{Louis Rose}, editors: {\sl
  \bibinfo{booktitle}{{TTC} 2011: Fifth Transformation Tool Contest, Z\"urich,
  Switzerland, June 29-30 2011}}.

\bibitemdeclare{inproceedings}{gretl-icmt2011}
\bibitem{gretl-icmt2011}
\bibinfo{author}{Tassilo Horn} \& \bibinfo{author}{J\"urgen Ebert}
  (\bibinfo{year}{2011}): \emph{\bibinfo{title}{{The GReTL Transformation
  Language}}}.
\newblock In \bibinfo{editor}{Jordi Cabot} \& \bibinfo{editor}{Eelco Visser},
  editors: {\sl \bibinfo{booktitle}{Theory and Practice of Model
  Transformations, Fourth International Conference, ICMT 2011, Zurich,
  Switzerland, June 27-28, 2011. Proceedings}}, {\sl \bibinfo{series}{Lecture
  Notes in Computer Science}} \bibinfo{volume}{6707},
  \bibinfo{publisher}{Springer}, pp. \bibinfo{pages}{183--197},
  \doi{10.1007/978-3-642-21732-6\_13}.

\end{thebibliography}

\end{document}